\documentclass
[aps,pra,amsfonts,amssymb,twocolumn,amsmath,preprintnumbers,nofootinbib,floatfix,
showpacs]{revtex4-1}%
\usepackage[dvips]{graphics}
\usepackage{graphicx}
\usepackage{bm}
\usepackage{amsmath}
\usepackage{amsfonts}
\usepackage{amssymb}
\usepackage{xcolor}
\usepackage{subfigure}
\usepackage{hyperref,hypcap}
\usepackage{braket}
\usepackage{commath}%
\providecommand{\U}[1]{\protect\rule{.1in}{.1in}}

\begin{document}

\title{Adiabatically Induced Orbital Magnetization}
\author{Cong Xiao}
\affiliation{Department of Physics, The University of Texas at Austin, Austin, Texas 78712, USA}
\author{Yafei Ren}
\affiliation{Department of Physics, The University of Texas at Austin, Austin, Texas 78712, USA}
\author{Bangguo Xiong}
\affiliation{Department of Physics, The University of Texas at Austin, Austin, Texas 78712, USA}

\begin{abstract}
A semiclassical theory for the orbital magnetization due to adiabatic evolutions of
Bloch electronic states is proposed. It renders a unified theory for the periodic-evolution
pumped orbital magnetization and the orbital
magnetoelectric response in insulators by revealing that these two phenomena
are the only instances where the induced magnetization is gauge invariant.
This theory also accounts for the electric-field induced intrinsic orbital
magnetization in two-dimensional metals and Chern insulators.
We illustrate the orbital magnetization pumped by microscopic local rotations of atoms, which
correspond to phonon modes with angular momentum, in toy models based on
honeycomb lattice, and the results are comparable to the pumped spin
magnetization via strong Rashba spin orbit coupling. We also show the vital
role of the orbital magnetoelectricity in validating the Mott relation between
the intrinsic nonlinear anomalous Hall and Ettingshausen effects.
\end{abstract}
\maketitle


\section{Introduction}

Understanding the orbital magnetization in crystalline solids is among the
most important objectives of orbitronics \cite{Zhang2005,Go2017,Bhowal2020}.
Unlike its spin companion, the orbital magnetization of Bloch electrons in the
absence of external perturbations is already hard to access quantum
mechanically, as it does not correspond to a bounded operator. This
nonlocality is finally accounted for by a Berry phase formula
\cite{Xiao2005,Resta2005,Shi2007} that gives significantly distinct orbital
magnetization from the atom-centered approximation when combined with
\textit{ab-initio} calculations in various magnetic materials
\cite{Ceresoli2010,Lopez2012,Hanke2016}. In the presence of external driving
electric fields, the extrinsic responses of spin and orbital magnetization,
namely the spin and orbital Edelstein effects, are given similarly by the
magnetic moments averaged over current-carrying states
\cite{Edelstein1990,Murakami2015,Mak2017,Pesin2018,Salemi2019,Mertig2020}. In
contrast, the intrinsic responses of them, i.e., the spin and orbital
magnetoelectric effects, are completely different due to the nonlocal nature
of the magnetic dipole operator. Specifically, the spin magnetoelectricity
\cite{Garate2009,Garate2010} is dictated by a Berry curvature following the
ubiquitous character of intrinsic linear responses of a local operator
\cite{Dong2020}, while the orbital one consists of a Chern-Simons three form
and a perturbative term of the reciprocal-space Berry connection
\cite{Moore2010,Vanderbilt2010,Lee2011,Gao2014}. Besides, the magnetization
pumped by periodic adiabatic processes in band insulators has been studied
recently by a density matrix approach evaluating the time-averaged expectation
value of the spin and magnetic dipole operators \cite{Luka2019,Murakami2020}.
In this approach, the orbital magnetization can only be obtained in the
Wannier basis \cite{Luka2019}, in contrast to the spin one that can be
evaluated in the Bloch representation.

Up to date, the orbital magnetization induced by electric fields and by
periodic adiabatic processes are treated by different theories. Whether both
phenomena have a deep connection and if they can emerge in a unified
theoretical framework are still unknown. In this work, we develop a
semiclassical theory for the magnetization induced by adiabatic
evolutions of Bloch electronic states. In general, the adiabatically induced
orbital magnetization is gauge dependent due to the presence of the electric
current in the second Chern form of Berry curvatures. Noticeably, the orbital
magnetoelectric effect and the periodic-evolution pumped orbital magnetization
emerge as the only instances where the induced magnetization is
gauge invariant due to the elimination of its explicit time dependence. Our work
thus renders a unified theory of both phenomena in insulators with vanishing
Chern numbers. Besides, unlike the Chern-Simons contribution deduced from the
second Chern form current, the induced magnetization due to the perturbed
Berry connection is well defined irrespective of Chern invariants and of
insulators or metals. As a result, the orbital magnetoelectricity in
two-dimensional (2D) metals and Chern insulators, which had long been
hard to approach, is also attained in our theory.

We apply our theory to illustrate the orbital magnetization pumped by
microscopic atomic rotations, which correspond to phonon modes with angular
momentum \cite{Zhang2014,Garanin2015}, in toy models based on the honeycomb
lattice. The results are comparable to the pumped spin magnetization via a
strong Rashba spin orbit coupling \cite{Murakami2020}. We also show the vital
role played by the electric-field induced orbital magnetization in the
nonlinear intrinsic anomalous Ettingshausen effect in 2D metallic systems. In
particular, the Mott relation is validated in intrinsic nonlinear transport by
subtracting the magnetization component of the thermal current in the second
order of the electric field.

This paper is organized as follows. In Sec. II we lay out the semiclassical
theory of Bloch electrons, which is employed to study the adiabatically
induced orbital magnetization in metals and insulators, respectively, in Sec.
III and Sec. IV. A case study of the orbital magnetization pumped by local
rotations of atoms and the application of the orbital magnetoelectricity to
the intrinsic nonlinear anomalous Ettingshausen effect are shown in Sec. V,
followed by a summary in Sec. VI. Some technical details of the theory are
presented in Appendices A and B for the convenience of interested readers.

\section{Semiclassical theory}

In the semiclassical description, the Hamiltonian felt by a narrow wave packet
centered around position $\boldsymbol{r}_{c}$ is $\hat{H}=\hat{H}_{c}+\hat
{H}^{\prime}$ in the first order gradient expansion, where $\hat{H}_{c}%
=\hat{H}_{0}\left(  \boldsymbol{\hat{r}},\boldsymbol{\hat{p}};w\left(
\boldsymbol{r}_{c}\right)  ,R\left(  t\right)  \right)  +\sum_{\alpha
}\boldsymbol{\hat{\theta}}^{\left(  \alpha\right)  }\cdot\boldsymbol{h}%
^{\left(  \alpha\right)  }\left(  \boldsymbol{r}_{c},t\right)  \ $is the local
Hamiltonian and $\hat{H}^{\prime}=\frac{1}{2}\{\left(  \hat{r}-r_{c}\right)
_{i},\frac{\partial\hat{H}_{c}}{\partial r_{ci}}\}$ is the gradient correction
\cite{Sundaram1999}. The Einstein summation convention is implied for repeated
Cartesian indices $i$, $j$ and $s$ henceforth. $\hat{H}_{0}$ is the local
approximation of the genuine Hamiltonian. The most general $\hat{H}_{0}$
considered here includes $w\left(  \boldsymbol{r}_{c}\right)  $ that
represents possible nonuniform static mechanical fields varying slowly on the
scale of the wave packet as well as $R\left(  t\right)  $ serving as a
parameter whose time evolution is adiabatic. Besides, in order to implement
the variational approach to obtain the local density of a bounded observable,
we add the auxiliary term $\sum_{\alpha}\boldsymbol{\hat{\theta}}^{\left(
\alpha\right)  }\cdot\boldsymbol{h}^{\left(  \alpha\right)  }\left(
\boldsymbol{r},t\right)  $ into the Hamiltonian and expand it around
$\boldsymbol{r}_{c}$. Here $\boldsymbol{\hat{\theta}}^{\left(  \alpha\right)
}$ ($\alpha=1,2,..$) is a set of bounded observable operators, each of which
is assumed to be a vector for simple notations, without losing generality.
$\boldsymbol{h}^{\left(  \alpha\right)  }\left(  \boldsymbol{r},t\right)  $
denotes the conjugate slowly varying external fields, and $\boldsymbol{h}%
^{\left(  \alpha\right)  }\left(  \boldsymbol{r}_{c},t\right)  $ changes
adiabatically in the parameter space $\left(  t,\boldsymbol{r}_{c}\right)  $.
At the end of the calculation the auxiliary term is set to zero, i.e.,
$\boldsymbol{h}^{\left(  \alpha\right)  }=0$.

In the semiclassical theory that is accurate to the first order of spatial
gradients and of time derivative \cite{Sundaram1999,Xiao2010}, the wave packet
$|W_{n}\left(  \boldsymbol{k},\boldsymbol{r}_{c},t\right)  \rangle
=\sum_{\boldsymbol{p}}C_{n\boldsymbol{p}}|\psi_{n\boldsymbol{p}}\rangle$ is
constructed by superposing the local Bloch states $|\psi_{n\boldsymbol{p}%
}\left(  \boldsymbol{r}_{c},t\right)  \rangle=e^{i\boldsymbol{p}%
\cdot\boldsymbol{\hat{r}}}|u_{n\boldsymbol{p}}\left(  \boldsymbol{r}%
_{c},t\right)  \rangle$ of $\hat{H}_{c}$. Here $n$ and $\hbar\boldsymbol{p}$
are the band index and crystal momentum, respectively, and the coefficient
$C_{n\boldsymbol{p}}$ is sharply distributed around the wave vector
$\boldsymbol{k}$ of the wave packet, obeying $\sum_{\boldsymbol{p}}\left\vert
C_{n\boldsymbol{p}}\right\vert ^{2}=1$. To simplify notations, all the band
quantities without explicit band index are considered for band $n$, unless
otherwise noted. Throughout this study we consider nondegenerate bands to
simplify the analysis and assume they are so separated that adiabatic
evolutions are feasible. The wave packet Lagrangian reads (set $\hbar=1$)
\begin{equation}
L=\langle W|i\frac{d}{dt}-\hat{H}|W\rangle=\boldsymbol{\dot{r}}_{c}%
\cdot\boldsymbol{k}+\boldsymbol{\dot{k}}\cdot\boldsymbol{\mathcal{A}}%
^{k}+\boldsymbol{\dot{r}}_{c}\cdot\boldsymbol{\mathcal{A}}^{r_{c}}%
+\mathcal{A}^{t}-\tilde{\varepsilon},\label{L}%
\end{equation}
where $\boldsymbol{\mathcal{A}}^{k/r_{c}}=\langle u_{n}|i\partial
_{\boldsymbol{k/r}_{c}}\mathbf{|}u_{n}\rangle$\ and $\mathcal{A}^{t}=\langle
u_{n}|i\partial_{t}\mathbf{|}u_{n}\rangle$ are the Berry connections derived
from the periodic part $|u_{n}\left(  \boldsymbol{k},\boldsymbol{r}%
_{c},t\right)  \rangle$ of the Bloch wave \cite{Sundaram1999}. The
noncanonical form of the Lagrangian due to Berry connections implies that
$\left(  \boldsymbol{r}_{c},\boldsymbol{k}\right)  $\ are not canonical
variables, thus the measure of the phase space spanned by $\left(
\boldsymbol{r}_{c},\boldsymbol{k}\right)  $ should be modified, with the
result \cite{Xiao2005} $\mathcal{D}=1+\Omega_{ii}^{kr_{c}}$. Here $\Omega
_{ii}^{kr_{c}}$ is the trace of the Berry curvature $\Omega_{ij}^{kr_{c}%
}\equiv\partial_{k_{i}}\mathcal{A}_{j}^{r_{c}}-\partial_{r_{cj}}%
\mathcal{A}_{i}^{k}$, and other Berry curvatures are formed similarly. The
wave-packet energy is given by $\tilde{\varepsilon}=\varepsilon+\delta
\varepsilon$ up to first order gradients, where $\varepsilon_{n}\left(
\boldsymbol{k},\boldsymbol{r}_{c},t\right)  $ is the local Bloch energy,
$\delta\varepsilon_{n}=\operatorname{Re}\sum_{n_{1}\neq n}%
\boldsymbol{\mathcal{A}}_{nn_{1}}^{k}\cdot(\partial_{\boldsymbol{r}_{c}}%
\hat{H}_{c})_{n_{1}n}$ and $\boldsymbol{\mathcal{A}}_{nn_{1}}^{k}=\langle
u_{n}|i\partial_{\boldsymbol{k}}\mathbf{|}u_{n_{1}}\rangle$ is the interband
Berry connection.

When going beyond the above first-order theory, the wave packet is no longer
dictated only by the Bloch states of local Hamiltonian $\hat{H}_{c}$ but is
modified by $\hat{H}^{\prime}$ up to the linear order of spatial gradients
\cite{Gao2014}. In the well established second-order theory \cite{Gao2019} the
inhomogeneity appears only in electromagnetic gauge potentials, whereas the
following results account for weak inhomogeneities of mechanical fields
conjugate to general bounded operators. This generalization is necessary to
obtain the adiabatically induced orbital magnetization carried by Bloch
electrons by calculating the magnetization current, which is manifested only
in nonuniform systems \cite{Cooper1997,Xiao2020EM}. For this purpose, it is
sufficient to retain results up to the order of the product spatial and time derivatives.

In the second-order theory the wave packet takes the form of \cite{Gao2019}
\begin{equation}
|W_{n}\left(  \boldsymbol{k},\boldsymbol{r}_{c},t\right)  \rangle
=\sum_{\boldsymbol{p}}e^{i\boldsymbol{p}\cdot\boldsymbol{\hat{r}}%
}[C_{n\boldsymbol{p}}|u_{n\boldsymbol{p}}\rangle+\sum_{n_{1}\neq n}%
C_{n_{1}\boldsymbol{p}}|u_{n_{1}\boldsymbol{p}}\rangle],
\end{equation}
where (derivations presented in Appendix A)
\begin{equation}
C_{n_{1}}=\frac{\Delta_{n_{1}n}C_{n}}{\varepsilon_{n\boldsymbol{p}%
}-\varepsilon_{n_{1}\boldsymbol{p}}}-i\boldsymbol{\mathcal{A}}_{n_{1}n}%
^{r_{c}}\cdot(i\partial_{\boldsymbol{p}}+\boldsymbol{\mathcal{A}}_{n}%
^{p}-\boldsymbol{r}_{c})C_{n}\label{C}%
\end{equation}
incorporates spatial-gradient induced corrections. Here
\begin{align}
\Delta_{n_{1}n} &  \equiv\frac{i}{2}(\partial_{\boldsymbol{\hat{p}}}%
\cdot\partial_{\boldsymbol{r}_{c}}\hat{H}_{c})_{n_{1}n}+\sum_{n_{2}\neq
n}(\partial_{\boldsymbol{r}_{c}}\hat{H}_{c})_{n_{1}n_{2}}\cdot
\boldsymbol{\mathcal{A}}_{n_{2}n}^{p}\nonumber\\
&  -\boldsymbol{\mathcal{A}}_{n_{1}n}^{r_{c}}\cdot\boldsymbol{v}%
_{n}\label{delta}%
\end{align}
has the dimension of energy, with $\boldsymbol{\mathcal{A}}_{nn_{1}}^{r_{c}%
}=\langle u_{n}|i\partial_{\boldsymbol{r}_{c}}\mathbf{|}u_{n_{1}}\rangle$. The
wave-packet center appearing in Eq. (\ref{C}) is determined by $\boldsymbol{r}%
_{c}=\langle W|\boldsymbol{\hat{r}}|W\rangle=\partial_{\boldsymbol{k}}%
\gamma(\boldsymbol{k})+\boldsymbol{\mathcal{A}}^{k}+\boldsymbol{a}^{k}$, where
$-\gamma(\boldsymbol{p})$ is the phase of $C_{n\boldsymbol{p}}$. While the
first two terms appear already in the first order theory \cite{Sundaram1999},
the third term $\boldsymbol{a}_{n}^{k}\equiv2\operatorname{Re}\sum
_{\boldsymbol{p}}\sum_{n_{1}\neq n}C_{n\boldsymbol{p}}^{\ast}C_{n_{1}%
\boldsymbol{p}}\boldsymbol{\mathcal{A}}_{nn_{1}}^{p}$ is the inhomogeneity
induced positional shift of the wave-packet center. Gathering the above
results we get a gauge invariant expression
\begin{equation}
\boldsymbol{a}_{n}^{k}=2\operatorname{Re}\sum_{n_{1}\neq n}\frac
{\boldsymbol{\mathcal{A}}_{nn_{1}}^{k}\Delta_{n_{1}n}}{\varepsilon
_{n}-\varepsilon_{n_{1}}}-\partial_{\boldsymbol{k}}\cdot\mathcal{G}_{n}%
^{r_{c}k},\label{ak}%
\end{equation}
with $\mathcal{G}_{n}^{r_{c}k}=\operatorname{Re}\sum_{n_{1}\neq n}%
\boldsymbol{\mathcal{A}}_{nn_{1}}^{r_{c}}\boldsymbol{\mathcal{A}}_{n_{1}n}%
^{k}$ being the quantum metric tensor in $\left(  \boldsymbol{r}%
_{c},\boldsymbol{k}\right)  $ space.

One can then find that the wave-packet Lagrangian takes the same form [Eq.
(\ref{L})] as in the first order theory, but the involved Berry connections
are modified by inhomogeneity, i.e., $\boldsymbol{\mathcal{\tilde{A}}}%
^{k}=\boldsymbol{\mathcal{A}}^{k}+\boldsymbol{a}^{k}$ and $\mathcal{\tilde{A}%
}^{t}=\mathcal{A}^{t}+a^{t}$. The corrected Berry connections in fact take
similar structures, e.g.,%
\begin{equation}
a_{n}^{t}=2\operatorname{Re}\sum_{n_{1}\neq n}\frac{\mathcal{A}_{nn_{1}}%
^{t}\Delta_{n_{1}n}}{\varepsilon_{n}-\varepsilon_{n_{1}}}-\partial
_{\boldsymbol{k}}\cdot\boldsymbol{\mathcal{G}}_{n}^{r_{c}t},\label{at}%
\end{equation}
where $\boldsymbol{\mathcal{G}}_{n}^{r_{c}t}=\operatorname{Re}\sum_{n_{1}\neq
n}\boldsymbol{\mathcal{A}}_{nn_{1}}^{r_{c}}\mathcal{A}_{n_{1}n}^{t}$, and are
gauge invariant. Note that the correction to $\boldsymbol{\mathcal{A}}^{r_{c}%
}$ is not needed, since $\boldsymbol{\mathcal{A}}^{r_{c}}$ is already in the
first order of spatial gradients. Meanwhile, the wave-packet energy does not
receive further corrections at the order of the product spatial and time derivatives.

Having identified the wave-packet Lagrangian and the concomitant action $S$,
one gets directly the semiclassical dynamics of Bloch electrons following from
the Euler-Lagrange equation in $\left(  \boldsymbol{r}_{c},\boldsymbol{k}%
\right)  $ space. Furthermore, one can consider the local density of a bounded
observable $\boldsymbol{\hat{\theta}}$, of which the conjugate external field
is marked by $\boldsymbol{h}$, contributed by a Bloch-electron ensemble with
the occupation function $f_{n}\left(  \boldsymbol{r}_{c},\boldsymbol{k}%
,t\right)  $. The general recipe for this has been given recently as
\cite{Dong2020}
\begin{equation}
\boldsymbol{\theta}\left(  \boldsymbol{r},t\right)  =-\int\left[
d\boldsymbol{k}\right]  d\boldsymbol{r}_{c}\mathcal{D}f\frac{\delta S}%
{\delta\boldsymbol{h}\left(  \boldsymbol{r},t\right)  }|_{\boldsymbol{h}%
\rightarrow0},\label{Key}%
\end{equation}
where $\left[  d\boldsymbol{k}\right]  \equiv\sum_{n}d\boldsymbol{k}%
/(2\pi)^{d}$ with $d$ as the spatial dimensionality. In what follows we
suppress the notation $\boldsymbol{h}\rightarrow0$ but the results for various
adiabatic responses are calculated in this limit. We take $f(\tilde
{\varepsilon})$ as the Fermi distribution in order to focus on adiabatic
intrinsic contributions determined solely by band structures.

The spin magnetoelectricity and spin pumping by periodic adiabatic processes
can be readily obtained from the above formula in the first order of time
derivative, as detailed in Appendix B, while the orbital counterparts can only
be acquired through calculating the electric current, which is much more
involved and is elaborated in the next two sections.

\section{Orbital magnetization in metals}

\subsection{Nonlinear electric current}

In order to address the orbital magnetization induced by adiabatic evolutions,
we need to formulate the local charge current density up to the order of the
product spatial and time derivatives. To achieve this Eq. (\ref{Key}) is
considered in the case of $\boldsymbol{\hat{\theta}}=e\boldsymbol{\hat{v}}$
being the charge current operator, hence $\boldsymbol{h}=-\boldsymbol{A}$ is
the electromagnetic vector potential with a minus sign, which enters through
the minimal coupling, resulting in the chain rule $\partial_{-\boldsymbol{A}%
}=e\partial_{\boldsymbol{k}}$. We still use $\boldsymbol{k}$ to denote the
gauge invariant crystal momentum. In the following the center label $c$ is
suppressed, unless otherwise noted. After some manipulations, as shown in
Appendix B, we arrive at (hereafter $\int$ without integral variable is
shorthand for $\int\left[  d\boldsymbol{k}\right]  $, $f^{0}=f\left(
\varepsilon\right)  $)
\begin{align}
\boldsymbol{j}\left(  \boldsymbol{r},t\right)   &  =\boldsymbol{\nabla}%
\times(\boldsymbol{M}^{0}+\int f^{0}\partial_{\boldsymbol{B}}a^{t}%
)+\partial_{t}\int f^{0}e\boldsymbol{a}^{k}\nonumber\\
&  -e\int f^{0}(\Omega^{\boldsymbol{k}t}+\Omega_{s}^{k\left[  kr\right]
t}\boldsymbol{\hat{e}}_{s})\nonumber\\
&  -e\int\partial_{\varepsilon}f^{0}\delta\varepsilon\Omega^{\boldsymbol{k}%
t}+e\int(\partial_{\boldsymbol{k}}f^{0}a^{t}-\partial_{t}f^{0}\boldsymbol{a}%
^{k}),\label{general}%
\end{align}
where the second Chern form of the Berry curvature \cite{Xiao2009,Zhou2013} is
labeled as
\begin{equation}
\Omega_{s}^{k\left[  kr\right]  t}\equiv\Omega_{si}^{kk}\Omega_{i}^{rt}%
+\Omega_{si}^{kr}\Omega_{i}^{tk}+\Omega_{s}^{kt}\Omega_{ii}^{kr},
\end{equation}
$\boldsymbol{\hat{e}}_{s}$ is the unit vector in the $s$ direction, and
\begin{equation}
\boldsymbol{M}^{0}=\int(f^{0}\boldsymbol{m}+eg^{0}\boldsymbol{\Omega
}).\label{M-eq}%
\end{equation}
Here $\boldsymbol{m}_{n}=\frac{e}{2}\sum_{n_{1}\neq n}\boldsymbol{\mathcal{A}%
}_{nn_{1}}^{k}\times\boldsymbol{v}_{n_{1}n}$ is the orbital moment of a Bloch
electron, $\boldsymbol{\Omega}$ is the vector form of Berry curvature
$\Omega_{ij}^{kk}$, and $g^{0}=-\int_{\varepsilon}^{\infty}f(\eta)d\eta$ is
the grand potential density contributed by a Bloch electron. With the
symmetric gauge for the uniform magnetic field, Eq. (\ref{at}) gives
\begin{equation}
\partial_{\boldsymbol{B}}a^{t}=2\operatorname{Re}\sum_{n_{1}\neq n}%
\frac{-\mathcal{A}_{nn_{1}}^{t}\boldsymbol{m}_{n_{1}n}}{\varepsilon
_{n}-\varepsilon_{n_{1}}}+\frac{e}{2}\partial_{\boldsymbol{k}}\times
\boldsymbol{\mathcal{G}}_{n}^{kt},\label{at-1}%
\end{equation}
where $\boldsymbol{\mathcal{G}}_{n}^{kt}=\operatorname{Re}\sum_{n_{1}\neq
n}\boldsymbol{\mathcal{A}}_{nn_{1}}^{k}\mathcal{A}_{n_{1}n}^{t}$ is the
quantum metric in $\left(  \boldsymbol{k},t\right)  $ space, and
$\boldsymbol{m}_{n_{1}n}=-\partial_{\boldsymbol{B}}\Delta_{n_{1}n}$, which
will be elaborated later in combination with a more specific physical context.

Equation (\ref{general}) is the pivotal result of this paper. The first line
is of total spatial and time derivatives, hence is certainly intimately
related to the orbital magnetization and electric polarization. Apparently,
$\boldsymbol{M}^{0}$ is the magnetization that relies solely on instantaneous
electronic states and corresponds to the equilibrium orbital magnetization in
the static case \cite{Shi2007}. On the other hand, the magnetization and
polarization may not be determined by the first line of Eq. (\ref{general})
alone, as the second line can be relevant as well. This line consists of first
and second Chern forms of Berry curvatures, which underline various electronic
topological responses of insulators \cite{Xiao2009,Xiao2010,Qi2008,Essin2009}.
Besides, the last line signifies intrinsic Fermi-surface contributions to the
charge current density in metals, which are beyond the conventional Boltzmann
transport picture of conductors \cite{Ziman} and are distinct from intrinsic
Fermi-sea contributions to linear response.

Now we are in a position to compare Eq. (\ref{general}) with existing results
at the same order. The second line of this equation has been formulated in
inhomogeneous insulators \cite{Xiao2009,Qi2015} and metals \cite{Hayata2017}.
The specific case where the inhomogeneity enters only through the magnetic
vector potential has been studied in insulators with degenerate bands
\cite{Moore2010,Qi2008} and in metals \cite{Moore2016}. Meanwhile, these
pioneering studies disregarded the magnetization current in the first line of
Eq. (\ref{general}), especially the orbital magnetization induced by the Berry
connection $a^{t}$ due to adiabatic time evolutions [Eq. (\ref{at-1})].
However, there is an important physical context: orbital magnetoelectricity in
2D metals, which is contributed entirely by this gauge invariant term and
hence is beyond the scope of the aforementioned theories. We discuss this
subject shortly.

\subsection{Orbital magnetoelectricity in 2D}

To address the orbital magnetoelectricity, we consider the case that the
adiabatic time dependence stems entirely from the vector potential, i.e.,
$\boldsymbol{E}=-\partial_{t}\boldsymbol{A}$, with $\boldsymbol{E}$ being a
weak constant electric field, then $\partial_{t}=e\boldsymbol{E}\cdot
\partial_{\boldsymbol{k}}$, $a^{t}=\boldsymbol{a}^{k}\cdot e\boldsymbol{E}$
and $\Omega_{s}^{k\left[  kr\right]  t}=\Omega_{sj}^{k\left[  kr\right]
k}eE_{j}$. Thus the local charge current density [Eq. (\ref{general})] reduces
to%
\begin{align}
\boldsymbol{j}\left(  \boldsymbol{r}\right)   &  =\boldsymbol{\nabla}%
\times\lbrack\boldsymbol{M}^{0}+\int f^{0}\partial_{\boldsymbol{B}%
}(e\boldsymbol{E\cdot a}^{k})]\nonumber\\
&  -e^{2}\boldsymbol{E}\times\int f^{0}\boldsymbol{\Omega}-e^{2}E_{i}\int
f^{0}\Omega_{si}^{k\left[  kr\right]  k}\boldsymbol{\hat{e}}_{s}\nonumber\\
&  +e^{2}\boldsymbol{E}\times\int\partial_{\varepsilon}f^{0}(\boldsymbol{v}%
\times\boldsymbol{a}^{k}-\delta\varepsilon\boldsymbol{\Omega}%
).\label{j-static}%
\end{align}

To understand this current we first inspect the case when the spatial
dependence originates only from the vector potential, i.e., $\boldsymbol{B}%
=\boldsymbol{\nabla}\times\boldsymbol{A}$. Then it is apparent that%
\begin{equation}
\boldsymbol{j}=-e^{2}\boldsymbol{E}\times\int\{f^{0}\boldsymbol{\Omega
}-\partial_{\varepsilon}f^{0}[\boldsymbol{v}\times(\boldsymbol{a}%
^{k})^{\boldsymbol{B}}+\left(  \boldsymbol{m}\cdot\boldsymbol{B}\right)
\boldsymbol{\Omega}]\},\label{NLH}%
\end{equation}
where $(\boldsymbol{a}^{k})^{\boldsymbol{B}}$\ is proportional to the magnetic
field. This result recovers the intrinsic magneto-nonlinear Hall current of
order $EB$ that was obtained previously by a different method \cite{Gao2014}.

On the other hand, to identify the orbital magnetization one could introduce
the spatial dependence from other inhomogeneous external mechanical fields. By
doing so one may expect that the local current density in bulk can be
decomposed into a transport and a magnetization component, namely
\cite{Cooper1997}%
\begin{equation}
\boldsymbol{j}=\boldsymbol{j}^{\text{tr}}+\boldsymbol{\nabla}\times
\boldsymbol{M},\label{decomposition}%
\end{equation}
where the transport current $\boldsymbol{j}^{\text{tr}}$ contributes to the
net flow through the sample. In 2D the second Chern form current is enforced
to vanish due to $\Omega_{si}^{k\left[  kr\right]  k}=0$, hence Eq.
(\ref{decomposition}) is rescued with $\boldsymbol{j}^{\text{tr}}$ taking the
same form as the above magneto-nonlinear Hall current Eq. (\ref{NLH}) and%
\begin{equation}
\boldsymbol{M}=\boldsymbol{M}^{0}+\int f^{0}\partial_{\boldsymbol{B}%
}(e\boldsymbol{E\cdot a}^{k}).\label{E-induced M}%
\end{equation}
Recall that $\boldsymbol{a}^{k}$ is physically a positional shift of a
semiclassical Bloch electron, the electric work upon this shift implies
immediately an orbital magnetization. This electric-field induced
magnetization is in agreement with what is obtained recently by a different
method \cite{Xiao2020OM}, but the present derivation is much simpler, even
though starting from a more generic framework. In 2D, $\boldsymbol{M}$ is a
pseudoscalar and is well defined irrespective of metals or Chern insulators.

Owing to the gauge invariance of $\boldsymbol{a}^{k}$, it is legitimate to
define the orbital magnetoelectric susceptibility contributed by each Bloch
electron in 2D $\alpha_{ij}^{\text{o}}$ via
\begin{equation}
\partial M_{j}/\partial E_{i}=\int f^{0}\alpha_{ij}^{\text{o}}. \label{2D OM}%
\end{equation}
$\alpha_{ij}^{\text{o}}$ takes a gauge invariant form ($j=z$ in 2D)%
\begin{equation}
\alpha_{ij}^{\text{o}}=-2e\operatorname{Re}\sum_{n_{1}\neq n}\frac
{(\mathcal{A}_{i}^{k})_{nn_{1}}(m_{j})_{n_{1}n}}{\varepsilon_{n}%
-\varepsilon_{n_{1}}}+\frac{e^{2}}{2}\epsilon_{jls}\partial_{k_{l}%
}(\mathcal{G}_{si}^{kk})_{n}, \label{OMP}%
\end{equation}
with $\mathcal{G}_{n}^{kk}=\operatorname{Re}\sum_{n_{1}\neq n}%
\boldsymbol{\mathcal{A}}_{nn_{1}}^{k}\boldsymbol{\mathcal{A}}_{n_{1}n}^{k}$ as
the $\boldsymbol{k}$-space quantum metric \cite{QM2020}.

It is interesting to compare $\alpha_{ij}^{\text{o}}$ with the spin
magnetoelectric susceptibility contributed by each Bloch electron, which is
given by the first term of Eq. (\ref{OMP}) with $\boldsymbol{m}_{n_{1}n}$
replaced by the interband elements of spin magnetic moment. Since
$\boldsymbol{m}_{n_{1}n}=-\frac{e}{2}\sum_{n_{2}\neq n}(\boldsymbol{v}%
_{n_{1}n_{2}}+\delta_{n_{2}n_{1}}\boldsymbol{v}_{n})\times
\boldsymbol{\mathcal{A}}_{n_{2}n}^{k}$ reduces to the familiar orbital moment
$\boldsymbol{m}_{n}$ when $n_{1}=n$, it can be deemed as an interband orbital
magnetic moment. Despite this similarity between spin and orbital
magnetoelectric susceptibility, the distinction is apparent: the
$\boldsymbol{k}$-space dipole moment of the quantum metric $\partial_{k_{l}%
}\mathcal{G}_{si}^{kk}$ does not have a counterpart in spin
magnetoelectricity. Noticeably, for two-band metallic systems with
particle-hole symmetry, the first term of $\alpha_{ij}^{\text{o}}$ vanishes,
hence $\alpha_{ij}^{\text{o}}=\frac{e^{2}}{2}\epsilon_{jls}\partial_{k_{l}%
}\mathcal{G}_{si}^{kk}$ is given solely by the quantum metric dipole, which is
an intrinsic Fermi surface effect.

Before closing this section, we note that in 3D insulators with nonvanishing
$\boldsymbol{k}$-space\ Chern invariants or 3D metals, the second Chern form
current in Eq. (\ref{j-static}) obviously poses a difficulty in pursuing a
gauge invariant decomposition in the form of Eq. (\ref{decomposition}). This
difficulty raises the question as to whether the electric-field induced
orbital magnetization can be defined as a bulk quantity in such systems. At
the present stage this is still an open question
\cite{Chen2012,Bergman2011,Qi2011} and is left for future efforts.

\section{Orbital magnetization in non-Chern insulators}

Now we turn to the nonlinear electric current in insulators in the general
case of adiabatic time evolutions and spatial dependence, under the assumption
of vanishing Chern numbers in all the pertinent parameter spaces. Great
simplifications of Eq. (\ref{general}) occur in insulators. First, the
Fermi-surface terms vanish and the Fermi-sea ones are contributed by fully
occupied bands. Then, according to the antisymmetric decomposition of the
second Chern form%
\begin{equation}
\Omega_{s}^{k\left[  kr\right]  t}=\partial_{k_{s}}CS_{ii}^{tkr}%
-\partial_{k_{i}}CS_{si}^{tkr}-\partial_{r_{i}}CS_{is}^{tkk}-\partial
_{t}CS_{sii}^{kkr}, \label{CS}%
\end{equation}
where the involved Chern-Simons\ three forms read, e.g., $CS_{si}^{tkr}%
=\frac{1}{2}(\mathcal{A}^{t}\Omega_{si}^{kr}+\mathcal{A}_{s}^{k}\Omega
_{i}^{rt}+\mathcal{A}_{i}^{r}\Omega_{s}^{tk})$, $CS_{is}^{tkk}=\frac{1}%
{2}(\mathcal{A}^{t}\Omega_{is}^{kk}+\mathcal{A}_{i}^{k}\Omega_{s}%
^{kt}+\mathcal{A}_{s}^{k}\Omega_{i}^{tk})$ and $CS_{sii}^{kkr}=\frac{1}%
{2}(\mathcal{A}_{s}^{k}\Omega_{ii}^{kr}+\mathcal{A}_{i}^{k}\Omega_{is}%
^{rk}+\mathcal{A}_{i}^{r}\Omega_{si}^{kk})$, the current density takes the
form of%
\begin{equation}
\boldsymbol{j}\left(  \boldsymbol{r},t\right)  =\boldsymbol{\nabla}%
\times\boldsymbol{M}\left(  \boldsymbol{r},t\right)  +\partial_{t}%
\boldsymbol{P}\left(  \boldsymbol{r},t\right)  . \label{current-insulator}%
\end{equation}
Here we have taken the $\boldsymbol{k}$-space periodic gauge for Bloch wave
functions, and
\begin{align}
\boldsymbol{M}  &  =\boldsymbol{M}^{0}+\int(\partial_{\boldsymbol{B}}%
a^{t}-e\frac{1}{2}\epsilon_{lis}CS_{is}^{tkk}\boldsymbol{\hat{e}}%
_{l}),\label{M}\\
\boldsymbol{P}  &  =e\int(\boldsymbol{\mathcal{A}}^{k}+\boldsymbol{a}%
^{k}+CS_{sii}^{kkr}\boldsymbol{\hat{e}}_{s}) \label{P}%
\end{align}
can be deemed as the orbital magnetization and polarization induced,
respectively, by the adiabatic time evolution and spatial inhomogeneity.

One can tell from Eq. (\ref{general}) that the perturbative contribution
$\partial_{\boldsymbol{B}}a^{t}$ to the orbital magnetization is well defined
regardless of Chern numbers in $\left(  \boldsymbol{k},t\right)  $ space and
is invariant under a gauge transformation of Bloch wave functions (a phase
transformation is compatible with the $\boldsymbol{k}$-space periodic gauge).
In contrast, the Chern-Simons orbital magnetization deduced from the second
Chern form current is only well defined in insulators with vanishing $\left(
\boldsymbol{k},t\right)  $-space Chern numbers. It changes under the gauge
transformation. It can be readily shown that this gauge dependence is
permitted by the inherent degrees of freedom of $\boldsymbol{M}\left(
\boldsymbol{r},t\right)  $ and $\boldsymbol{P}\left(  \boldsymbol{r},t\right)
$ determined by the invariance of the local current density Eq.
(\ref{current-insulator}) \cite{Hirst1997} (e.g., in the 2D case the inherent
degrees of freedom of $\boldsymbol{M}$ and $\boldsymbol{P}$ are $M_{z}%
\boldsymbol{\hat{z}}\rightarrow M_{z}\boldsymbol{\hat{z}}-\partial_{t}%
\chi\boldsymbol{\hat{z}}\ $and$\ \boldsymbol{P}\rightarrow\boldsymbol{P}%
+\boldsymbol{\nabla}\times(\chi\boldsymbol{\hat{z}})$, with a scalar field
$\chi(\boldsymbol{r},t)$).

This gauge dependence also implies, on the other hand, the necessity of
removing the time dependence of the orbital magnetization and the spatial
dependence of the electric polarization if one would like to pursue gauge
invariant definitions of them. Therefore, there are generally two ways to have
a gauge invariant orbital magnetization: to either eliminate the explicit time
dependence of $\boldsymbol{M}$ or pursue the definition upon an average over
time. These two approaches correspond to two important physical contexts --
orbital magnetoelectric response and orbital magnetization pumping --\ that
are addressed separately in the following two subsections.

\subsection{Orbital magnetoelectric response}

When the time and spatial dependence concerns only the electromagnetic gauge
potentials, the explicit time dependence of $\boldsymbol{M}$ and spatial
dependence of $\boldsymbol{P}$ are removed due to the minimal coupling. This
is the case of the orbital magnetoelectric response in insulators, which
includes two dual effects: a constant electric (magnetic) field induces an
orbital magnetization (electric polarization). Most previous derivations are
designed for only one of the two dual effects
\cite{Moore2010,Vanderbilt2010,Gao2014,Lee2011}, while a theory capable of
both simultaneously is rare \cite{Sipe2020}. Here we show that they are
readily derived from the present theory.

On the one hand, when the time dependence appears solely as $\boldsymbol{E}%
=-\partial_{t}\boldsymbol{A}$, one has $\int CS_{is}^{tkk}=\frac{-\theta
e}{4\pi^{2}}\epsilon_{isj}E_{j}$, where $\theta=-\int\frac{d^{3}k}{4\pi
}\boldsymbol{\mathcal{A}}^{k}\cdot\boldsymbol{\Omega}$ is the abelian version
of the so called $\theta$-term \cite{Xiao2010,Essin2009}. Then $\boldsymbol{M}%
$ [Eq. (\ref{M})] becomes a time-independent orbital magnetization%
\begin{equation}
\boldsymbol{M}=\boldsymbol{M}^{0}+\partial_{\boldsymbol{B}}\int\boldsymbol{a}%
^{k}\cdot e\boldsymbol{E}+\frac{e^{2}}{4\pi^{2}}\theta\boldsymbol{E}.
\label{OME}%
\end{equation}
On the other hand, when the spatial dependence appears only as a magnetic
field, $\int CS_{sii}^{kkr}=\frac{\theta e}{4\pi^{2}}B_{s}$. Thus one can
identify $\boldsymbol{P}$ as a uniform polarization, which is in agreement
with the previous theory \cite{Gao2014}, and verify $\partial M_{i}/\partial
E_{j}=\partial P_{j}/\partial B_{i}$.

\subsection{Periodic-evolution pumped orbital magnetization}

It is also possible to define, based on $\boldsymbol{M}$ and $\boldsymbol{P}$,
the time averaged orbital magnetization in time periodic systems and spatially
averaged polarization in spatially periodic systems. Here we concentrate on
the magnetization, and the polarization can be discussed similarly. If the
adiabatic time dependence of the electronic Hamiltonian is periodic with
period $T$ and the Chern invariants in $\left(  \boldsymbol{k},t\right)  $
space are zero, then the time averaged $\boldsymbol{M}$ is gauge invariant and
can be perceived as the orbital magnetization pumped by the periodic
evolution, namely
\begin{equation}
\boldsymbol{\bar{M}}\left(  \boldsymbol{r}\right)  =\int_{0}^{T}\frac{dt}%
{T}\boldsymbol{M}\left(  \boldsymbol{r},t\right)  .
\end{equation}
We use the notation $\boldsymbol{\bar{M}}$ to distinguish from the
instantaneous magnetization $\boldsymbol{M}\left(  \boldsymbol{r},t\right)  $
in Eq. (\ref{current-insulator}). In 2D the $\boldsymbol{\nabla}\phi$ degree
of freedom of the magnetization is irrelevant and the so defined
$\boldsymbol{\bar{M}}\left(  \boldsymbol{r}\right)  $ gives the orbital
magnetization unambiguously. In both 2D and 3D this $\boldsymbol{\bar{M}}%
$\ coincides with the abelian version of the so-called geometric orbital
magnetization obtained by a density matrix approach evaluating the
time-averaged expectation value of the magnetic dipole operator in the Wannier
basis in homogeneous band insulators \cite{Luka2019}. On the other hand, the
present theory does not invoke the Wannier basis and accounts also for weak
inhomogeneous systems.

\section{Applications}

\subsection{Model illustration of orbital magnetization pumped by local
rotations of atoms}

To illustrate the above theory, we consider a minimal model for the orbital
magnetization due to the periodic adiabatic evolution of electronic states
induced by microscopic local rotations of atoms. Such a model is not required
to possess the spin-orbit coupling, in contrast to the spin magnetization
induced by local circulations of atoms that is only possible with the aid of
spin-orbit coupling \cite{Murakami2020}. The minimal spatial dimensionality
for rotational motions is two, and the model should have a gap. Moreover, the
second Chern form current can be nonzero only if the dimension of the
Hamiltonian is larger than two \cite{Xiao2009}. Therefore, we here consider a
two-band model hence focus exclusively on the contribution
\begin{equation}
\boldsymbol{\bar{M}}=\int_{0}^{T}\frac{dt}{T}\int\frac{d\boldsymbol{k}}%
{(2\pi)^{2}}\partial_{\boldsymbol{B}}a^{t}\label{OM pumping}%
\end{equation}
from the perturbed Berry connection. According to the expression for
$\partial_{\boldsymbol{B}}a^{t}$ (only the first term of Eq. (\ref{at-1})
matters in insulators), one can easily verify that it can be nonvanishing in a
two-band model\ only if the particle-hole symmetry is broken.

\begin{figure}[ptb]
\includegraphics[width=9 cm]{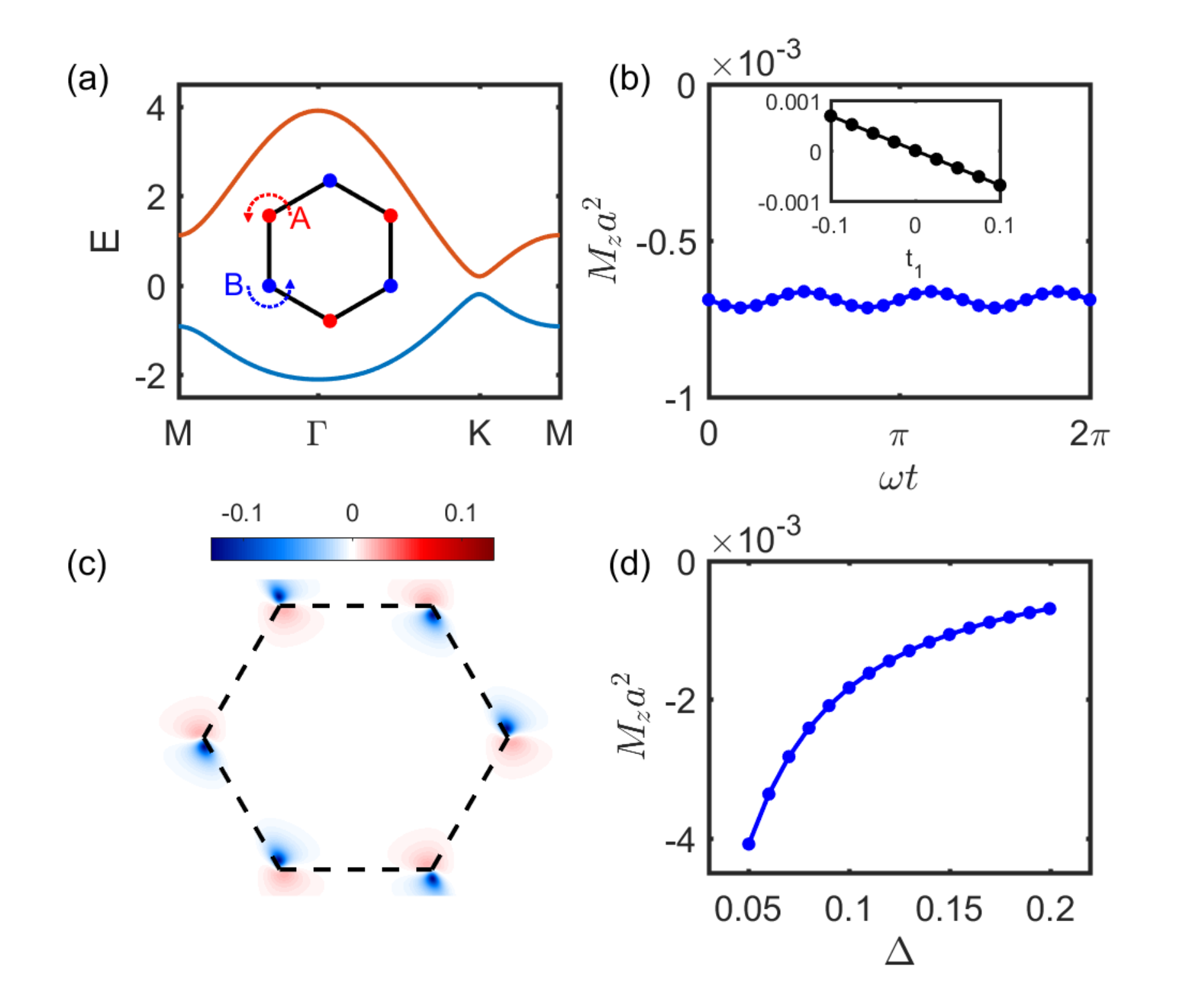}\caption{Model
illustration of the orbital magnetization of Bloch electrons induced by the
microscopic local rotation of atoms. (a) Band structure of a spinless-graphene
toy model with first and second nearest neighbor hoppings. (b) Time dependence
of the adiabatically induced magnetic moment in units of $e\omega a^{2}$ per
unit cell. Here $\omega$ is the angular frequency of the atomic rotation, $a$
is the lattice constant, and we take the parameters as $\Delta=0.2t_{0}$,
$t_{1}=0.1t_{0}$ and $\delta t_{0}=0.1t_{0}$. The insert shows that the pumped
orbital magnetization in one period of the local rotations of atoms is
proportional to the next nearest neighbor hopping parameter. (c)
$\boldsymbol{k}$-space\ distribution of $\partial_{\boldsymbol{B}}a^{t}$ of
the valence band. (d) Gap dependence of the pumped orbital magnetization.}%
\label{MurakamiSpinless}%
\end{figure}

Such a minimal model can thus be chosen as a spinless graphene-type one based
on the honeycomb lattice taking into account the next nearest neighbor
hopping, which is described by the Hamiltonian
\begin{equation}
\hat{H}(\boldsymbol{k})=t_{0}(F_{\mathrm{R}}\sigma_{x}-F_{\mathrm{Im}}%
\sigma_{y})+\Delta\sigma_{z}+t_{1}F_{\mathrm{NN}}\sigma_{0},
\end{equation}
where $F_{\mathrm{R}}=2\cos x\cos y+\cos2y$, $F_{\mathrm{Im}}=2\cos x\sin
y-\sin2y$, and $F_{\mathrm{NN}}=2\cos(2x/\sqrt{3})+4\cos(x/\sqrt{3})\cos
\sqrt{3}y+3$ with $x=k_{x}a_{0}\sqrt{3}/2$, $y=k_{y}a_{0}/2$ and $a_{0}$ being
the inter-atomic distance. The first and second nearest neighbor hoppings are
$t_{0}$ and $t_{1}$, respectively, and a nonzero $t_{1}$ breaks the
particle-hole symmetry. The staggered sublattice potential strength is
$\Delta$.

Next, we add an adiabatic perturbation term due to the microscopic local
rotation of atoms, and mainly follow the treatment introduced in Ref.
\cite{Murakami2020}, where a right handed circularly polarized optical phonon
mode at $\Gamma$ point is considered, with frequency $\omega$ and displacement
vectors%
\begin{equation}
\boldsymbol{u}_{A}=u_{0}(\cos\omega t,\sin\omega t),\text{ \ }\boldsymbol{u}%
_{B}=-\boldsymbol{u}_{A}%
\end{equation}
of A and B atoms on the two sublattices. There is a phase difference $\pi$
between the circular rotations of atoms A and B (see also Fig
\ref{MurakamiSpinless}(a)), thus the nearest neighbor bond lengths change with
time by the microscopic local rotations, while the next nearest neighbor ones
do not. One can hence take these rotations as the modulation of the nearest
neighbor hopping. By writing down the tight-binding Hamiltonian and converting
it to a $\boldsymbol{k}$-space one, the resultant adiabatic perturbation to
$\hat{H}(\boldsymbol{k})$ reads
\begin{align}
\delta\hat{H}(\boldsymbol{k},t)= &  -\delta t_{0}(\sigma_{x}\sin y+\sigma
_{y}\cos y)\sqrt{3}\sin x\cos\omega t\nonumber\\
+ &  \delta t_{0}[(\cos x\cos y-\cos2y)\sigma_{x}\nonumber\\
&  -(\cos x\sin y+\sin2y)\sigma_{y}]\sin\omega t,
\end{align}
where $\delta t_{0}\propto2u_{0}$ arises from the change of the first nearest
neighbor hopping energy due to the variation of the inter-atomic distance by
the local rotations \cite{Murakami2020}. In our calculation, we consider the
chemical potential inside the band gap and set $t_{0}$ as the energy unit,
$t_{1}=0.1t_{0}$, and $\delta t_{0}=0.1t_{0}$.

With $\Delta=0.2t_{0}$, we plot energy bands in the absence of phonons in
Fig.~\ref{MurakamiSpinless}(a) where a band gap opens and the energy bands do
not show particle-hole symmetry. In the presence of phonons, the adiabatic
evolution of the electronic states due to the local rotations of A and B atoms
leads to a time-dependent orbital magnetic moment, which is plotted in
Fig.~\ref{MurakamiSpinless}(b) in units of $\frac{et_{0}}{\hbar}a^{2}%
\frac{\hbar\omega}{t_{0}}$ (magnetic moment upon an area of $a^{2}$) in an
evolution period. It is apparent that the induced orbital magnetization is
proportional to the phonon frequency $\omega$. By using the parameters of
graphene with $t_{0}=3$ eV and the lattice constant $a=\sqrt{3}a_{0}=2.46$
\r{A}, $\frac{et_{0}}{\hbar}a^{2}$ is about 4.77$\mu_{B}$ with $\mu_{B}$ as
the Bohr magneton. One can find a weak oscillation with a nonzero net
contribution (about $-3.5\times10^{-3}$ $\mu_{B}\frac{\hbar\omega}{t_{0}}$)
upon one period of the local rotations of atoms, which implies a nonzero
pumping of orbital magnetization. The $\boldsymbol{k}$-resolved instantaneous
orbital magnetization $\partial_{\boldsymbol{B}}a^{t}$ is plotted in
Fig.~\ref{MurakamiSpinless}(c) where one can find that the main contribution
is from K and K' points with minimal interband spacings. As the band gap
decreases, the magnitude of the induced magnetization increases rapidly
(Fig.~\ref{MurakamiSpinless}(d)).

We verify that the induced orbital magnetization vanishes with the next
nearest neighbor hopping parameter $t_{1}$ in the insert in
Fig.~\ref{MurakamiSpinless}(b), which also shows that the pumped magnetization
is proportional to $t_{1}$ at least up to $t_{1}=0.1t_{0}$. The resultant
orbital magnetization pumping in the above toy model is of the same order as
the pumped spin magnetization in a honeycomb lattice with very strong Rashba
spin-orbit coupling (the Rashba coefficient equals to $0.4t_{0}$)
\cite{Murakami2020}. We also mention that, in this latter four-band model,
breaking the particle-hole symmetry is not required for supporting nonzero
orbital magnetization pumped by rotations of atoms, and the pumped orbital and
spin magnetization are also generally comparable (not shown here).
Furthermore, taking $\hbar\omega/t_{0}=0.1$ (the ratio of the typical energy
scales of phonons and electrons), the magnetic-moment pumping due to the
periodic adiabatic change of electronic states induced by microscopic
rotations of atoms is of the order of the nuclear magneton.

\subsection{Intrinsic nonlinear anomalous Ettingshausen effect in 2D metals}

Not only the electric current but also a thermal current carried by Bloch
electrons can be induced by an applied electric field. In the intrinsic linear
thermal current response to the electric field, the zero-field orbital
magnetization plays a vital role \cite{Cooper1997,Xiao2006,Xiao2020EM}. It is
therefore anticipated that the electric-field induced orbital magnetization is
indispensable in the second-order nonlinear intrinsic thermal current response
to the electric field, i.e., the nonlinear intrinsic anomalous Ettingshausen
effect. In this subsection we discuss in more detail the semiclassical picture
of the electric-field induced orbital magnetization in 2D metals, and point
out its key role in the intrinsic nonlinear anomalous Ettingshausen effect.

First, the electric-field modified orbital magnetization can be recast into an
instructive form%
\begin{equation}
\boldsymbol{M}=\int(f^{0}\boldsymbol{\tilde{m}}+eg^{0}\boldsymbol{\tilde
{\Omega}})\label{M guess}%
\end{equation}
in analogy to the magnetization (\ref{M-eq}) in the absence of electric
fields. Here $\boldsymbol{\tilde{\Omega}}=\partial_{\boldsymbol{k}}%
\times\lbrack\boldsymbol{\mathcal{A}}^{k}+(\boldsymbol{a}^{k})^{\boldsymbol{E}%
}]$ is the electric-field modified Berry curvature, and%
\begin{equation}
\boldsymbol{\tilde{m}}=\boldsymbol{m}+\partial_{\boldsymbol{B}}%
(e\boldsymbol{E}\cdot\boldsymbol{a}^{k})+e\boldsymbol{v}^{0}\times
(\boldsymbol{a}^{k})^{\boldsymbol{E}}%
\end{equation}
is the orbital moment $\boldsymbol{\tilde{m}}=\frac{e}{2}\langle W|\left(
\boldsymbol{\hat{r}}-\boldsymbol{r}_{c}\right)  \times\boldsymbol{\hat{v}%
}|W\rangle$ up to the first order of the electric field. $\boldsymbol{m}$ is
the zero-field orbital moment, $\boldsymbol{v}^{0}$ is the band velocity, and
$(\boldsymbol{a}^{k})^{\boldsymbol{E}}$ is the positional shift linear in the
electric field \cite{Gao2014}.

Second, a coarse graining process based on the wave-packet description of
Bloch electrons \cite{Xiao2006} shows that, the electric-field induced local
energy current density up to the second order is given by%
\begin{equation}
\boldsymbol{j}^{\text{E}}=-e\boldsymbol{E}\times\int f^{0}\varepsilon
\boldsymbol{\tilde{\Omega}}-\boldsymbol{E}\times\int f^{0}\boldsymbol{\tilde
{m}}.
\end{equation}
A magnetization current $\boldsymbol{j}^{\text{E,mag}}$ should be discounted
to obtain the transport energy current density \cite{Cooper1997}
$\boldsymbol{j}^{\text{E,tr}}=\boldsymbol{j}^{\text{E}}-\boldsymbol{j}%
^{\text{E,mag}}$. In uniform crystals, the energy magnetization current at the
linear order of the electric field is given by the the material-dependent part
of the Poynting vector describing the energy flow \cite{Xiao2006}:
$\boldsymbol{j}^{\text{E,mag}}=-\boldsymbol{E}\times\boldsymbol{M}^{0}$. In
the present nonlinear response, one has
\begin{equation}
\boldsymbol{j}^{\text{E,mag}}=-\boldsymbol{E}\times\boldsymbol{M}.
\end{equation}
Consequently, the transport thermal current is given by
\begin{equation}
\boldsymbol{j}^{\text{h,tr}}=\boldsymbol{j}^{\text{E,tr}}-\frac{\mu}%
{e}\boldsymbol{j}=-e\boldsymbol{E}\times T\int s\left(  \varepsilon\right)
\boldsymbol{\tilde{\Omega}},
\end{equation}
where $s\left(  \varepsilon\right)  =\left[  \left(  \varepsilon-\mu\right)
f^{0}-g^{0}\right]  /T$ is the entropy density contributed by a particular
Bloch state, $\mu$ is the chemical potential, $T$ is the temperature, and we
have made use of the result for the intrinsic nonlinear Hall electric current
\cite{Gao2014} $\boldsymbol{j}=-e^{2}\boldsymbol{E}\times\int f^{0}%
\boldsymbol{\tilde{\Omega}}$.

By integration by parts, the entropy density takes the form of $s\left(
\varepsilon\right)  =\int d\eta\left(  \eta-\mu\right)  \partial_{\mu}f\left(
\eta\right)  \theta\left(  \eta-\varepsilon\right)  /T$, which renders the
thermal transport current to be%
\begin{equation}
\frac{\boldsymbol{j}^{\text{h,tr}}}{T}=-\frac{1}{e}\int d\eta\frac{\eta-\mu
}{T}\frac{\partial f\left(  \eta\right)  }{\partial\eta}\boldsymbol{j}\left(
\eta\right)  .
\end{equation}
Here $\boldsymbol{j}\left(  \eta\right)  =-e^{2}\boldsymbol{E}\times\int
\theta\left(  \eta-\varepsilon\right)  \boldsymbol{\tilde{\Omega}}$ is the
nonlinear Hall electric current at zero temperature with Fermi energy $\eta$.
This equation is completely parallel to the generalized Mott relation between
the transport thermal current and electric current in the linear order of
electric fields. At low temperatures much less than the distances between the
chemical potential and band edges, the Sommerfeld expansion is legitimate
\cite{Xiao2016}, hence the entropy density reduces to $s\left(  \varepsilon
\right)  =\frac{1}{3}\pi^{2}k_{B}^{2}T\delta\left(  \mu-\varepsilon\right)  $,
which is concentrated on the Fermi surface and decays dramatically away from
it. Then the standard form of the Mott relation follows%
\begin{equation}
\boldsymbol{j}^{\text{h}}/T=(\pi^{2}k_{B}^{2}T/3e)[\partial\boldsymbol{j}%
\left(  \varepsilon\right)  /\partial\varepsilon]|_{\varepsilon=\mu}.
\end{equation}
Therefore, we extend the regime of validity of the Mott relation to the
second-order intrinsic thermoelectric current responses to a constant electric field.

\section{Summary}

We have formulated a semiclassical theory for the orbital magnetization
induced by general adiabatic evolutions of Bloch electronic states. This
theory starts from formulating the electric current density in bulk, from
which the magnetization can be extracted. The induced orbital magnetization is
gauge dependent in general case but is gauge invariant only when the adiabatic
time dependence is implicit or averaged out. These two cases correspond to the
orbital magnetoelectric response and the periodic-evolution pumped orbital magnetization.

In the orbital magnetoelectric effect the adiabatic evolution is driven by a
constant electric field, and the time dependence is only implicit through the
evolution of mechanical crystal momentum. Thus the pertinent second Chern form
current vanishes in 2D, making the 2D orbital magnetoelectricity governed
completely by the perturbative term of the reciprocal-space Berry connection
(Eqs. (\ref{2D OM}) and (\ref{OMP})), irrespective of insulators or metals.
The role of the orbital magnetoelectricity in the nonlinear intrinsic
anomalous Ettingshausen effect, which is proposed here as a transverse thermal
current response in the second order of the driving electric field, has also
been revealed in 2D metals. On the other hand, the orbital magnetoelectricity
and the nonlinear intrinsic anomalous Ettingshausen effect in 3D metals are
beyond the scope of the present theory. They may not be determined solely by
bulk considerations and are left for future efforts.

In the context of the orbital magnetization pumped by periodic adiabatic
evolutions in non-Chern insulators, the Chern-Simons contribution deduced from
the second Chern form current can be present even in 2D. Meanwhile, as a
second Chern form can be nonzero only if the system has more than two bands
\cite{Xiao2009}, in a two-band minimal model the pumped magnetization is
dictated solely by the perturbative term of the time component of the Berry
connection [Eqs. (\ref{OM pumping}) and (\ref{at-1})]. We illustrated the
orbital magnetization pumping due to the periodic adiabatic change of
electronic states induced by microscopic rotations of atoms in toy models
based on the honeycomb lattice. The induced magnetization is of the same order
as the pumped spin magnetization via strong Rashba spin orbit coupling.

The presented formulation is based on the assumption of well separated
nondegenerate Bloch bands, whereas to explore the semiclassical theories in
the case of degenerate bands and of closely located bands with possible
non-adiabatic effects \cite{Tu2020,Woods2020} need separate studies.

\begin{acknowledgments}
We thank Luka Trifunovic and Liang Dong for enlightening discussions.
This work was supported by NSF (EFMA-1641101) and Welch Foundation (F-1255).
\end{acknowledgments}

\appendix

\section{Derivation of gradient corrected wave-packet state}

Given that the perturbative $\hat{H}^{\prime}$ is the form of the gradient
correction, the first-order wave-packet reads
\begin{align}
|W_{n}(\boldsymbol{k})\rangle &  =\sum_{\boldsymbol{p}}C_{n\boldsymbol{p}%
}\left[  |\psi_{n\boldsymbol{p}}\rangle\right.  \nonumber\\
&  \left.  +\sum_{n_{1}\boldsymbol{p}_{1}\neq n\boldsymbol{p}}\frac
{\langle\psi_{n_{1}\boldsymbol{p}_{1}}|\hat{H}^{\prime}|\psi_{n\boldsymbol{p}%
}\rangle}{\varepsilon_{n\boldsymbol{p}}-\varepsilon_{n_{1}\boldsymbol{p}_{1}}%
}|\psi_{n_{1}\boldsymbol{p}_{1}}\rangle\right]  \nonumber\\
&  =\sum_{\boldsymbol{p}}e^{i\boldsymbol{p}\cdot\boldsymbol{\hat{r}}%
}[C_{n\boldsymbol{p}}|u_{n\boldsymbol{p}}\rangle+\sum_{n_{1}\neq n}%
C_{n_{1}\boldsymbol{p}}|u_{n_{1}\boldsymbol{p}}\rangle],
\end{align}
where%
\begin{align}
C_{n_{1}\boldsymbol{p}} &  =\sum_{\boldsymbol{p}_{1}}C_{n\boldsymbol{p}_{1}%
}\frac{\langle\psi_{n_{1}\boldsymbol{p}}|\hat{H}^{\prime}|\psi
_{n\boldsymbol{p}_{1}}\rangle}{\varepsilon_{n\boldsymbol{p}_{1}}%
-\varepsilon_{n_{1}\boldsymbol{p}}}\nonumber\\
&  =\sum_{n_{2}\boldsymbol{p}_{1}}C_{n\boldsymbol{p}_{1}}\frac{(\partial
_{\boldsymbol{r}_{c}}\hat{H}_{c}\left(  \boldsymbol{p}\right)  )_{n_{1}n_{2}%
}\cdot i(\boldsymbol{D}_{\boldsymbol{p}})_{n_{2}n}\delta_{\boldsymbol{pp}_{1}%
}}{\varepsilon_{n\boldsymbol{p}_{1}}-\varepsilon_{n_{1}\boldsymbol{p}}%
}\nonumber\\
&  +\sum_{n_{2}\boldsymbol{p}_{1}}C_{n\boldsymbol{p}_{1}}\frac{(\partial
_{\boldsymbol{r}_{c}}\hat{H}_{c}\left(  \boldsymbol{p}_{1}\right)  )_{n_{2}%
n}\cdot i(\boldsymbol{D}_{\boldsymbol{p}})_{n_{1}n_{2}}\delta_{\boldsymbol{pp}%
_{1}}}{\varepsilon_{n\boldsymbol{p}_{1}}-\varepsilon_{n_{1}\boldsymbol{p}}}.
\end{align}
Here we introduced the notation
\begin{equation}
i(\boldsymbol{D}_{\boldsymbol{p}})_{n_{2}n}\equiv i\partial_{\boldsymbol{p}%
}\delta_{n_{2}n}+\boldsymbol{\mathcal{A}}_{n_{2}n}^{p}-\boldsymbol{r}%
_{c}\delta_{n_{2}n}.
\end{equation}
After some manipulations we get Eq. (\ref{C}) with%
\begin{align}
\Delta_{n_{1}n} &  =\frac{1}{2}(i\partial_{\boldsymbol{p}}%
+\boldsymbol{\mathcal{A}}_{n_{1}}^{p}-\boldsymbol{\mathcal{A}}_{n}^{p}%
)\cdot(\partial_{\boldsymbol{r}_{c}}\hat{H}_{c})_{n_{1}n}\nonumber\\
&  +\frac{1}{2}\sum_{n_{2}\neq n_{1}}\boldsymbol{\mathcal{A}}_{n_{1}n_{2}}%
^{p}\cdot(\partial_{\boldsymbol{r}_{c}}\hat{H}_{c})_{n_{2}n}\nonumber\\
&  +\frac{1}{2}\sum_{n_{2}\neq n}(\partial_{\boldsymbol{r}_{c}}\hat{H}%
_{c})_{n_{1}n_{2}}\boldsymbol{\mathcal{A}}_{n_{2}n}^{p}%
-\boldsymbol{\mathcal{A}}_{n_{1}n}^{r_{c}}\cdot\boldsymbol{v}_{n}.
\end{align}
By using
\begin{align}
\partial_{\boldsymbol{p}}\cdot(\partial_{\boldsymbol{r}_{c}}\hat{H}%
_{c})_{n_{1}n} &  =(\partial_{\boldsymbol{\hat{p}}}\cdot\partial
_{\boldsymbol{r}_{c}}\hat{H}_{c})_{n_{1}n}\nonumber\\
&  +i\sum_{n_{2}}[\boldsymbol{\mathcal{A}}_{n_{1}n_{2}}^{p}\cdot
(\partial_{\boldsymbol{r}_{c}}\hat{H}_{c})_{n_{2}n}\nonumber\\
&  -(\partial_{\boldsymbol{r}_{c}}\hat{H}_{c})_{n_{1}n_{2}}\cdot
\boldsymbol{\mathcal{A}}_{n_{2}n}^{p}],
\end{align}
one gets Eq. (\ref{delta}). Here we also note that Eqs. (\ref{C}) and
(\ref{delta}) has also been obtained by a different method in a recent
preprint \cite{Zhao2020}. In addition, when $[\partial_{\boldsymbol{r}_{c}%
}\hat{H}_{c},\boldsymbol{\hat{r}]}=0$, Eq. (\ref{delta}) reduces to
$\Delta_{n_{1}n}\equiv\sum_{n_{2}\neq n}(\partial_{\boldsymbol{r}_{c}}\hat
{H}_{c})_{n_{1}n_{2}}\cdot\boldsymbol{\mathcal{A}}_{n_{2}n}^{p}%
-\boldsymbol{\mathcal{A}}_{n_{1}n}^{r_{c}}\cdot\boldsymbol{v}_{n}$.

\bigskip

\section{Derivation of Eq. (\ref{general})}

As shown in Ref. \cite{Dong2020}, one has%
\begin{align}
\int\left[  d\boldsymbol{k}\right]  d\boldsymbol{r}_{c}\mathcal{D}%
f\frac{\delta S}{\delta\boldsymbol{h}}  &  =\int\mathcal{D}f[\frac{\partial
L}{\partial\boldsymbol{h}}-\frac{d}{dt}\frac{\partial L}{\partial(\partial
_{t}\boldsymbol{h})}]\nonumber\\
&  -\partial_{r_{i}}\int\mathcal{D}f[\frac{\partial L}{\partial(\partial
_{i}\boldsymbol{h})}-\frac{\partial L}{\partial(\partial_{t}\boldsymbol{h}%
)}\dot{r}_{i}],
\end{align}
then the field variation formula (\ref{Key}) yields%
\begin{equation}
\theta_{s}=\int\mathcal{D}f(\partial_{h_{s}}\tilde{\varepsilon}-\tilde{\Omega
}_{s}^{h\mathcal{T}})-\partial_{r_{i}}\int f[d_{is}^{\theta}-\frac{\partial
a^{t}}{\partial\left(  \partial_{i}h_{s}\right)  }].
\end{equation}
Here $\tilde{\Omega}_{s}^{h\mathcal{T}}=\Omega_{si}^{hk}\dot{k}_{i}%
+\Omega_{si}^{hr}\dot{r}_{i}+\tilde{\Omega}_{s}^{ht}$, where $\dot{k}%
_{i}=-\partial_{i}\varepsilon+\Omega_{i}^{rt}$ and $\dot{r}_{i}=\partial
_{k_{i}}\varepsilon-\Omega_{i}^{kt}$ up to the order of the product spatial
and time derivatives according to the equations of motion, and only the Berry
curvature $\Omega^{\boldsymbol{h}t}$ need be modified by inhomogeneity:%
\begin{equation}
\tilde{\Omega}^{ht}=\partial_{\boldsymbol{h}}\mathcal{\tilde{A}}^{t}%
-\partial_{t}\boldsymbol{\mathcal{\tilde{A}}}^{h},\text{ \ }%
\boldsymbol{\mathcal{\tilde{A}}}^{h}=\frac{\partial\mathcal{\tilde{A}}^{t}%
}{\partial(\partial_{t}\boldsymbol{h})}=\boldsymbol{\mathcal{A}}%
^{h}+\boldsymbol{a}^{h}.
\end{equation}
Then we arrive at%
\begin{align}
\theta_{s}\left(  \boldsymbol{r},t\right)   &  =\partial_{h_{s}}(G-\int
f^{0}a^{t})-\partial_{r_{i}}[D_{is}^{\theta}-\int f^{0}\frac{\partial a^{t}%
}{\partial\left(  \partial_{i}h_{s}\right)  }]\nonumber\\
&  -\int f^{0}(\Omega_{s}^{ht}+\Omega_{s}^{h\left[  kr\right]  t}%
)+\partial_{t}\int f^{0}a_{s}^{h}\nonumber\\
&  -\int\partial_{\varepsilon}f^{0}\delta\varepsilon\Omega_{s}^{ht}%
+\int(\partial_{h_{s}}f^{0}a^{t}-\partial_{t}f^{0}a_{s}^{h}), \label{main}%
\end{align}
where each term is gauge invariant, and the second Chern form of the Berry
curvature is labeled as%
\begin{equation}
\Omega_{si}^{hk}\Omega_{i}^{rt}+\Omega_{si}^{hr}\Omega_{i}^{tk}+\Omega
_{s}^{ht}\Omega_{ii}^{kr}\equiv\Omega_{s}^{h\left[  kr\right]  t}.
\end{equation}
Besides, $G=\int\mathcal{D}g\left(  \tilde{\varepsilon}\right)  $ is the
electronic grand potential density and is evaluated to the first order of
gradients, $g\left(  \tilde{\varepsilon}\right)  =g^{0}+f^{0}\delta
\varepsilon$, and $\partial g\left(  \tilde{\varepsilon}\right)
/\partial\tilde{\varepsilon}=f\left(  \tilde{\varepsilon}\right)
=f^{0}+\partial_{\varepsilon}f^{0}\delta\varepsilon$, with $g^{0}%
=-\int_{\varepsilon}^{\infty}f(\eta)d\eta$ and $f^{0}=f\left(  \varepsilon
\right)  $. $D_{is}^{\theta}=\int(f^{0}d_{is}^{\theta}+g^{0}\Omega_{is}^{kh})$
is the $\theta$-dipole density of the electron system, with $d_{ij}^{\theta
}=\partial\delta\varepsilon_{n}/\partial\left(  \partial_{i}h_{j}\right)  $
being the $\theta$-dipole moment of a semiclassical Bloch electron
\cite{Dong2020}.

In the case of $\boldsymbol{\hat{\theta}}=e\boldsymbol{\hat{v}}$ and
$\boldsymbol{h}=-\boldsymbol{A}$, Eq. (\ref{main}) reduces to Eq.
(\ref{general}).

On the other hand, in the absence of inhomogeneity Eq. (\ref{main}) reduces to
$\boldsymbol{\theta}\left(  t\right)  =\int f^{0}(\partial_{\boldsymbol{h}%
}\varepsilon-\boldsymbol{\Omega}^{\boldsymbol{h}t})$. In insulators, hence
zero temperature for electrons, one has
\begin{equation}
\boldsymbol{\theta}\left(  t\right)  =\int\langle u|\boldsymbol{\hat{\theta}%
}\mathbf{|}u\rangle-\int\boldsymbol{\Omega}^{\boldsymbol{h}t}. \label{bounded}%
\end{equation}
The first term on the right hand side is simply the average value of
$\boldsymbol{\theta}$ in the electron system obtained by using the
instantaneous Hamiltonian, whereas the second term is a geometric term related
to the Berry curvature in $\left(  t,\boldsymbol{h}\right)  $ space%
\begin{equation}
\boldsymbol{\Omega}^{\boldsymbol{h}t}=-\boldsymbol{\Omega}^{t\boldsymbol{h}%
}=2\operatorname{Re}\sum_{n_{1}\neq n}\frac{\mathcal{A}_{nn_{1}}%
^{t}\boldsymbol{\theta}_{n_{1}n}}{\varepsilon_{n}-\varepsilon_{n_{1}}}.
\label{Curvature}%
\end{equation}
Equation (\ref{bounded}) gives a unified account of diverse adiabatic
responses of bounded operators in band insulators, such as the spin
magnetoelectric effect, where the spin magnetization is induced by a weak
electric field, and the spin magnetization pumped by microscopic local
rotation of atoms \cite{Murakami2020}.


\begin{thebibliography}{99}                                                                                               %


\bibitem {Zhang2005}B. A. Bernevig, T. L. Hughes, and S.-C. Zhang, Phys. Rev.
Lett. \textbf{95}, 066601 (2005).

\bibitem {Go2017}D. Go, J.-P. Hanke, P. M. Buhl, F. Freimuth, G. Bihlmayer,
H.-W. Lee, Y. Mokrousov, and S. Blugel, Sci. Rep. \textbf{7}, 46742 (2017).

\bibitem {Bhowal2020}S. Bhowal and S. Satpathy, Phys. Rev. B \textbf{101},
121112(R) (2020).

\bibitem {Xiao2005}D. Xiao, J. Shi, and Q. Niu, Phys. Rev. Lett. \textbf{95},
137204 (2005).

\bibitem {Resta2005}T. Thonhauser, D. Ceresoli, D. Vanderbilt, and R. Resta,
Phys. Rev. Lett. \textbf{95}, 137205 (2005).

\bibitem {Shi2007}J. Shi, G. Vignale, D. Xiao, and Q. Niu, Phys. Rev. Lett.
\textbf{99}, 197202 (2007).

\bibitem {Ceresoli2010}D. Ceresoli, U. Gerstmann, A. P. Seitsonen, and F.
Mauri, Phys. Rev. B \textbf{81}, 060409(R) (2010).

\bibitem {Lopez2012}M. G. Lopez, D. Vanderbilt, T. Thonhauser, and I. Souza,
Phys. Rev. B \textbf{85}, 014435 (2012).

\bibitem {Hanke2016}J.-P. Hanke, F. Freimuth, A. K. Nandy, H. Zhang, S.
Blugel, and Y. Mokrousov, Phys. Rev. B \textbf{94}, 121114(R) (2016).

\bibitem {Edelstein1990}V. M. Edelstein, Solid State Commun. \textbf{73}, 233 (1990).

\bibitem {Murakami2015}T. Yoda, T. Yokoyama, and S. Murakami, Sci. Rep.
\textbf{5}, 12024 (2015).

\bibitem {Mak2017}J. Lee, Z. Wang, H. Xie, K. F. Mak, and J. Shan, Nat. Mater.
\textbf{16}, 887 (2017).

\bibitem {Pesin2018}C. \c{S}ahin, J. Rou, J. Ma, and D. A. Pesin, Phys. Rev. B
\textbf{97}, 205206 (2018).

\bibitem {Salemi2019}L. Salemi, M. Berritta, A. K. Nandy, and P. M. Oppeneer,
Nature Communications \textbf{10}, 5381 (2019).

\bibitem {Mertig2020}A. Johansson, B. Gobel, J. Henk, M. Bibes, and I. Mertig,
arXiv: 2006.14958

\bibitem {Garate2009}I. Garate and A. H. MacDonald, Phys. Rev. B \textbf{80},
134403 (2009).

\bibitem {Garate2010}I. Garate and M. Franz, Phys. Rev. Lett. \textbf{104},
146802 (2010).

\bibitem {Dong2020}L. Dong, C. Xiao, B. Xiong, and Q. Niu, Phys. Rev. Lett.
\textbf{124}, 066601 (2020).

\bibitem {Vanderbilt2010}A. Malashevich, I. Souza, S. Coh, and D. Vanderbilt,
New J. Phys. \textbf{12}, 053032 (2010).

\bibitem {Moore2010}A. M. Essin, A. M. Turner, J. E. Moore, and D. Vanderbilt,
Phys. Rev. B \textbf{81}, 205104 (2010).

\bibitem {Lee2011}K.-T. Chen and P. A. Lee, Phys. Rev. B \textbf{84}, 205137 (2011).

\bibitem {Gao2014}Y. Gao, S. A. Yang, and Q. Niu, Phys. Rev. Lett.
\textbf{112}, 166601 (2014).

\bibitem {Luka2019}L. Trifunovic, S. Ono, and H. Watanabe, Phys. Rev. B
\textbf{100}, 054408 (2019).

\bibitem {Murakami2020}M. Hamada and S. Murakami, Phys. Rev. Res. \textbf{2},
023275 (2020).

\bibitem {Zhang2014}L. Zhang and Q. Niu, Phys. Rev. Lett. \textbf{112}, 085503 (2014).

\bibitem {Garanin2015}D. A. Garanin and E. M. Chudnovsky, Phys. Rev. B
\textbf{92}, 024421 (2015).

\bibitem {Sundaram1999}G. Sundaram and Q. Niu, Phys. Rev. B \textbf{59}, 14915 (1999).

\bibitem {Xiao2010}D. Xiao, M.-C. Chang, and Q. Niu, Rev. Mod. Phys.
\textbf{82}, 1959 (2010).

\bibitem {Gao2019}Y. Gao, Frontiers of Physics \textbf{14}, 33404 (2019).

\bibitem {Cooper1997}N. R. Cooper, B. I. Halperin, and I. M. Ruzin, Phys. Rev.
B \textbf{55}, 2344 (1997).

\bibitem {Xiao2020EM}C. Xiao and Q. Niu, Phys. Rev. B \textbf{101, }235430 (2020).

\bibitem {Zhou2013}J.-H. Zhou, J. Hua, Q. Niu, and J.-R. Shi, Chin. Phys.
Lett. \textbf{30}, 027101 (2013).

\bibitem {Xiao2009}D. Xiao, J. Shi, D. P. Clougherty, and Q. Niu, Phys. Rev.
Lett. \textbf{102}, 087602 (2009).

\bibitem {Qi2008}X.-L. Qi, T. L. Hughes, and S.-C. Zhang, Phys. Rev. B
\textbf{78}, 195424 (2008).

\bibitem {Essin2009}A. M. Essin, J. E. Moore, and D. Vanderbilt, Phys. Rev.
Lett. \textbf{102}, 146805 (2009).

\bibitem {Ziman}J. M. Ziman, {\itshape}Principles of the Theory of Solids
(Cambridge University Press, Cambridge, 1972).

\bibitem {Qi2015}D. Bulmash, P. Hosur, S.-C. Zhang, and X.-L. Qi, Phys. Rev. X
\textbf{5}, 021018 (2015).

\bibitem {Hayata2017}T. Hayata and Y. Hidaka, Phys. Rev. B \textbf{95}, 125137 (2017).

\bibitem {Moore2016}D. Varjas, A. G. Grushin, R. Ilan, and J. E. Moore, Phys.
Rev. Lett. \textbf{117}, 257601 (2016).

\bibitem {Xiao2020OM}C. Xiao, H. Liu, J. Zhao, S. A. Yang, and Q. Niu, arXiv: 2002.01637

\bibitem {QM2020}A. Gianfrate, O. Bleu, L. Dominici, V. Ardizzone, M. De
Giorgi, D. Ballarini, G. Lerario, K. W. West, L. N. Pfeiffer, D. D.
Solnyshkov, D. Sanvitto, and G. Malpuech, Nature \textbf{578}, 381 (2020).

\bibitem {Chen2012}K.-T. Chen and P. A. Lee, Phys. Rev. B \textbf{86}, 195111 (2012).

\bibitem {Bergman2011}D. L. Bergman, Phys. Rev. Lett. \textbf{107}, 176801 (2011).

\bibitem {Qi2011}M. Barkeshli and X.-L. Qi, Phys. Rev. Lett. \textbf{107},
206602 (2011).

\bibitem {Hirst1997}L. L. Hirst, Rev. Mod. Phys. \textbf{69}, 607 (1997).

\bibitem {Sipe2020}P. T. Mahon and J. E. Sipe, Phys. Rev. Res. \textbf{2},
033126 (2020).

\bibitem {Xiao2006}D. Xiao, Y. Yao, Z. Fang, and Q. Niu, Phys. Rev. Lett.
\textbf{97}, 026603 (2006).

\bibitem {Xiao2016}C. Xiao, D. Li, and Z. Ma, Phys. Rev. B \textbf{93}, 075150 (2016).

\bibitem {Tu2020}Matisse Wei-Yuan Tu, C. Li, H. Yu, and W. Yao, Phys. Rev. B
\textbf{102}, 045423 (2020).

\bibitem {Woods2020}T. Stedman and L. M. Woods, Phys. Rev. Res. \textbf{2},
033086 (2020).

\bibitem {Zhao2020}Y. Zhao, Y. Gao, and D. Xiao, arXiv: 2009.09306
\end{thebibliography}
\end{document}